\documentclass[superscriptaddress,aps,prl,twocolumn,floatfix]{revtex4-2}

\usepackage{bm}
\usepackage{dcolumn}
\usepackage{graphicx}
\usepackage[colorlinks=true, linkcolor=blue, citecolor=blue, urlcolor=blue]{hyperref}
\usepackage{mathrsfs}
\usepackage{multirow}
\usepackage{rotating}
\usepackage{url}
\usepackage{color}
\usepackage{amsmath}
\usepackage{amssymb}

\graphicspath{{figures/}}
\begin{document}

\title{Emergent Interfacial Magnetism in Epitaxial RuO$_2$}

\author{Yudi Yang}
\thanks{These authors contributed equally to this work.}
\affiliation{Zhejiang University, Hangzhou, Zhejiang 310058, China}
\affiliation{Department of Physics, Westlake University, Hangzhou 310030, Zhejiang, China}
\affiliation{Institute of Natural Sciences, Westlake Institute for Advanced Study, Hangzhou 310024, Zhejiang, China}

\author{Zhuang Qian}
\thanks{These authors contributed equally to this work.}
\affiliation{Institute for Theoretical Sciences, Westlake Institute for Advanced Study, Westlake University, Hangzhou 310024, Zhejiang, China}

\author{Shi Liu}
\email{liushi@westlake.edu.cn}
\affiliation{Department of Physics, Westlake University, Hangzhou 310030, Zhejiang, China}
\affiliation{Institute of Natural Sciences, Westlake Institute for Advanced Study, Hangzhou 310024, Zhejiang, China}

\author{Congjun Wu}
\email{wucongjun@westlake.edu.cn}
\affiliation{Institute of Natural Sciences, Westlake Institute for Advanced Study, Hangzhou 310024, Zhejiang, China}
\affiliation{Institute for Theoretical Sciences, Westlake Institute for Advanced Study, Westlake University, Hangzhou 310024, Zhejiang, China}
\affiliation{New Cornerstone Science Laboratory, Department of Physics, Westlake University, Hangzhou 310024, Zhejiang, China}

\date{\today}

\begin{abstract}
The magnetic ground state of the altermagnet candidate RuO$_2$ remains controversial, with magnetic signatures observed mainly in epitaxial films. Here we show, using first-principles calculations, that magnetism in epitaxial RuO$_2$ can emerge as an interfacial boundary phase at TiO$_2$/RuO$_2$ interfaces. While TiO$_2$-induced epitaxial strain alone does not make (001)-oriented RuO$_2$ magnetic, explicit TiO$_2$/RuO$_2$ interfaces stabilize sizable Ru moments confined to the first few Ru layers. Charge-density and orbital-resolved analyses reveal interfacial electronic reconstruction, and substrate doping provides a route to tune the induced moments. In symmetric TiO$_2$/RuO$_2$/TiO$_2$ heterostructures, the two magnetic interfaces couple through the metallic RuO$_2$ spacer, producing a thickness-dependent alternation between weak-ferromagnetic and compensated altermagnetic states. Our results identify interface engineering as a practical route to stabilize and control fragile magnetism in RuO$_2$.
\end{abstract}

\maketitle
Altermagnetism combines the zero net magnetization of a compensated antiferromagnet with nonrelativistic momentum-dependent spin splitting, enabling spin-polarized transport without ferromagnetic stray fields \cite{Smejkal22p031042,Smejkal22p040501,Song25p473,Liu25p329}. In this state, opposite-spin sublattices are related by crystal rotations or mirrors rather than by translation or inversion. Among the earliest and most widely studied candidate altermagnets, metallic rutile RuO$_2$ has served as a representative platform~\cite{Smejkal20peaaz8809,Feng22p735,Rafael21p127701,Karube22p137201,Bai22p197202,Bai23p216701,Shao21p7061}. Yet its intrinsic magnetic ground state remains unsettled. Early neutron and resonant x-ray studies reported compensated magnetic order \cite{Berlijn17p077201,Zhu19p017202}, and subsequent thin-film experiments found transport signatures consistent with altermagnetism, including anomalous Hall responses, spin-splitting torques, spin-to-charge conversion, and nonlinear Hall effects~\cite{Feng22p735,Karube22p137201,Bai22p197202,Bai23p216701,Chu25p216703}.  In contrast, bulk-sensitive probes, including muon spin rotation, neutron diffraction, and spin-resolved photoemission, have found either no static magnetic order or no clear altermagnetic spin splitting in high-quality RuO$_2$ single crystals \cite{Hiraishi24p166702,Philipp24arxiv,Liu24p176401}.

This experimental dichotomy, recently reviewed by Li \textit{et al.} \cite{Li26p257001}, suggests that magnetic and altermagnetic signatures in RuO$_2$ are strongly sample dependent. They appear mainly in epitaxial films, whereas measurements on pristine bulk crystals increasingly support a nonmagnetic ground state. Additionally, many film-based signatures remain indirect and show appreciable sensitivity to sample quality, suggesting that extrinsic factors such as strain and defects may play an important role.

A natural explanation is that RuO$_2$ lies close to a magnetic instability. Spin-channel Fermi-liquid interactions can dynamically generate unconventional momentum-dependent spin order \cite{Wu04p036403,Wu07p115103}, and recent density-functional-theory (DFT) calculations suggest that RuO$_2$ is near a Landau--Pomeranchuk-type instability \cite{Qian25p174425}. In this near-critical regime, modest changes in hopping amplitudes, induced for example by epitaxial strain, or in Ru-$d$ occupation, modified by charge-carrier doping, can shift the balance among competing magnetic states.

This sensitivity is especially relevant for RuO$_2$ films grown on rutile TiO$_2$, whose structural compatibility enables coherent epitaxy along the commonly studied (001), (100), and (110) orientations \cite{Feng22p735,Li26p257001}. Because of the anisotropic lattice mismatch, TiO$_2$-induced strain has been regarded as a key control parameter. However, strain alone cannot fully account for the observations. DFT calculations for strained bulk RuO$_2$ show that TiO$_2$-imposed in-plane strain can stabilize magnetic states for the (100) and (110) orientations, but not for the (001)-oriented RuO$_2$ \cite{Forte25Strain}. This contrasts with the third-order nonlinear Hall response reported in 5-nm (001)-oriented RuO$_2$ films, which was attributed to intrinsic magnetic order \cite{Chu25p216703}. The discrepancy points to additional interfacial mechanisms beyond purely clamping-induced in-plane strain.

Here we show that the magnetism of epitaxial RuO$_2$ is not stabilized by strain alone, but emerges as an interfacial boundary phase driven by TiO$_2$-induced electronic reconstruction. Direct atomic contact with TiO$_2$ drives electron transfer into RuO$_2$; after electronic reconstruction, the added charge is preferentially accommodated by O orbitals, effectively hole-doping the Ru $d$ orbitals and nucleating sizable Ru moments confined to the boundary layers.  Additionally, these moments can be enhanced or suppressed by substrate charge doping. In symmetric TiO$_2$/RuO$_2$/TiO$_2$ heterostructures, the two magnetic interfaces remain coupled through the metallic RuO$_2$ spacer, with a thickness-dependent sign change reminiscent of Ruderman–Kittel–Kasuya–Yosida (RKKY)~\cite{Ruderman54p99,Kasuya56p45,Yosida57p893} exchange. This interfacial coupling enables deterministic selection between a fully compensated altermagnetic state and an uncompensated weak-ferromagnetic state, establishing a route to the on-demand engineering of fragile magnetism in near-critical RuO$_2$.

All DFT calculations were performed using \textsc{Quantum Espresso}~\cite{Giannozzi09p395502,Giannozzi17p465901} within the Perdew-Burke-Ernzerhof (PBE) form of the generalized-gradient approximation~\cite{Perdew96p3865} and optimized norm-conserving Vanderbilt pseudopotentials from the \textsc{PseudoDojo} library~\cite{Van18p39}, with a plane-wave cutoff of $80\,{\rm Ry}$. The Brillouin zone was sampled using $12\times12$ and $8\times8$ in-plane $k$-point meshes for the primitive and $\sqrt{2}\times\sqrt{2}$ in-plane cells, respectively, together with two out-of-plane $k$ points for periodic sandwich structures and one for slab geometries. We emphasize that no Hubbard $U$ correction was applied to the Ru $4d$ orbitals.

\begin{figure}[t]
  \centering
  \includegraphics[width=0.48\textwidth]{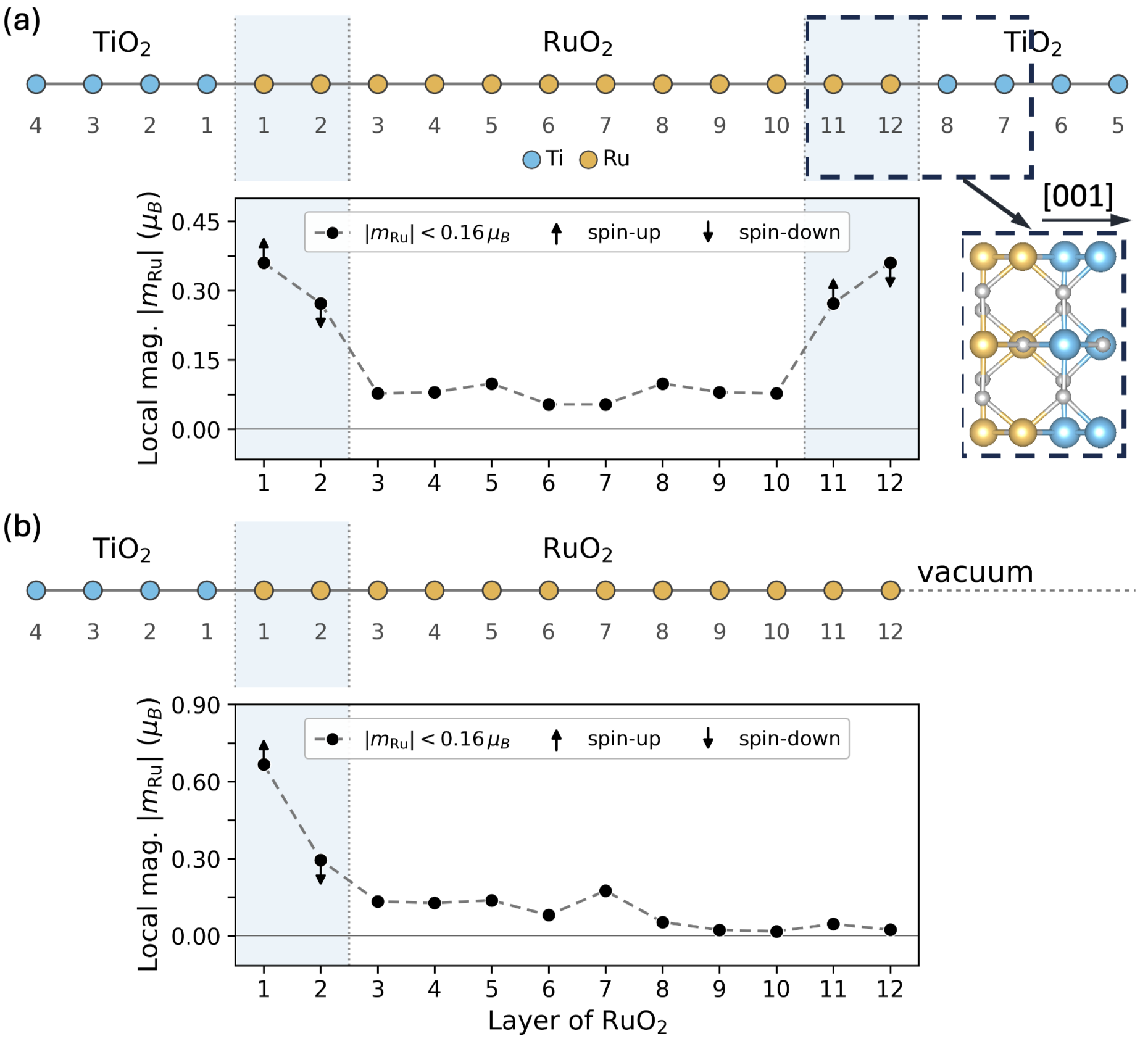}
  \caption{Layer-resolved local magnetic moments in (001)-oriented RuO$_2$/TiO$_2$ heterostructures.
(a) Top: schematic structural models, with blue and orange circles denoting Ti and Ru atomic layers, respectively. Bottom: spatial distribution of the absolute Ru local moments, $|m_{\mathrm{Ru}}|$, in a heterostructure where 12 RuO$_2$ layers are sandwiched between TiO$_2$ layers. The magnetic order is strongly localized in the RuO$_2$ boundary layers near both interfaces, indicated by the shaded regions, and decays rapidly toward the nonmagnetic interior. Local moments larger than $0.16\,\mu_B$ are marked by additional arrows indicating their spin orientations. The inset shows a detailed (001) view of the right-hand interface.
(b) Corresponding distribution in an asymmetric TiO$_2$/RuO$_2$/vacuum slab. The induced moments appear only near the direct TiO$_2$ contact and are absent at the vacuum-terminated surface.}
  \label{fig:interface}
\end{figure}

We first performed DFT calculations for bulk RuO$_2$ under various epitaxial strain conditions by fixing the in-plane lattice constants and allowing the out-of-plane lattice parameter and internal atomic coordinates to relax. 
Consistent with previous DFT calculations~\cite{Forte25Strain}, the in-plane strain imposed by TiO$_2$ alone is insufficient to make (001)-oriented RuO$_2$ magnetic (Fig.~S1). We then explicitly constructed TiO$_2$/RuO$_2$ heterostructures under periodic boundary conditions, corresponding to TiO$_2$/RuO$_2$ superlattices [see the atomic interface structures in Fig.~\ref{fig:interface}(a)]. In these calculations, the in-plane lattice constants of the (001) interface were fixed to those of TiO$_2$, while the out-of-plane lattice constant and all atomic positions were fully relaxed. Importantly, as shown in Fig.~\ref{fig:interface}(a), Ru atoms near the TiO$_2$/RuO$_2$ interfaces develop substantial local magnetic moments exceeding \(0.2\,\mu_{\rm B}\), and the interfacial moments adopt an uncompensated antiferromagnetic ordering. The moments are strongly localized near the interfaces and decay rapidly toward the interior of the RuO$_2$ slab, where only very small moments remain (below \(0.16\,\mu_{\rm B}\)). 

To further isolate the role of the TiO$_2$ interface, we constructed an asymmetric TiO$_2$/RuO$_2$/vacuum slab, in which RuO$_2$ has one interface with TiO$_2$ and one bare surface. Because periodic boundary conditions are applied in the in-plane directions, fixing the in-plane lattice vectors to those of TiO$_2$ ensures that the RuO$_2$ slab remains under the same epitaxial in-plane strain. As shown in Fig.~\ref{fig:interface}(b), sizable local magnetic moments again appear only in the RuO$_2$ layers closest to the TiO$_2$ interface, primarily within the first two Ru layers, and decay quickly away from it. In contrast, the Ru atoms near the vacuum-terminated surface remain essentially nonmagnetic. The absence of surface magnetism demonstrates that the magnetism cannot be attributed to epitaxial strain alone or to generic symmetry breaking. The interfacial electronic reconstruction at the TiO$_2$/RuO$_2$ interface is therefore the key ingredient responsible for the emergence of local magnetic moments. These results also show that RuO$_2$ can become magnetic at the interface in DFT, even without introducing a Hubbard \(U\) on the Ru \(4d\) orbitals. We note that additional calculations for the TiO$_2$/RuO$_2$ heterostructure with the in-plane lattice constants fixed to the bulk RuO$_2$ values show the absence of interfacial magnetism, indicating the concerted effect of in-plane strain and interfacial electronic reconstruction.

\begin{figure}[t]
  \centering
  \includegraphics[width=0.49\textwidth]{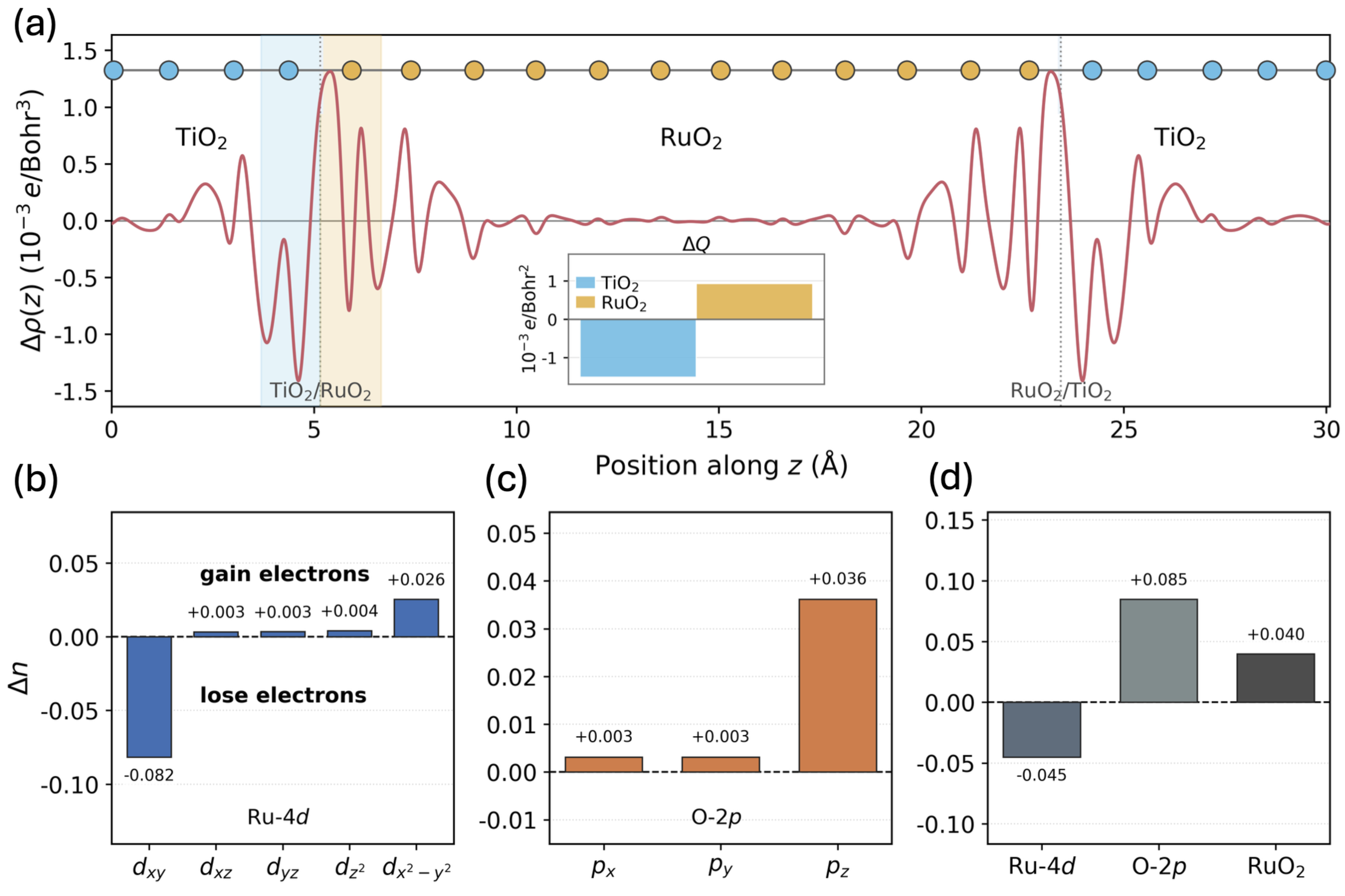}
  \caption{Interfacial electronic reconstruction in the TiO$_2$/RuO$_2$/TiO$_2$ heterostructure.
(a) Planar-averaged charge-density difference, $\Delta\rho(z)$, along the out-of-plane direction. The shaded regions denote the TiO$_2$- and RuO$_2$-side interface windows used to evaluate the signed integrated charge redistribution, $\Delta Q$, shown in the inset. The integrated charge indicates net electron transfer from TiO$_2$ to RuO$_2$ at the interface.
(b,c) Changes in orbital-projected L\"owdin occupations for one interfacial Ru atom and one neighboring O atom, respectively, referenced to the averages over the four central, bulk-like RuO$_2$ layers.
(d) Total occupation changes within one interfacial RuO$_2$ unit, consisting of one Ru atom and two O atoms. The total is computed in the projected subspace of the Ru $4d$ and O $2p$ orbitals. Positive values indicate the gain of electrons.}
  \label{fig:charge}
\end{figure}

To analyze charge transfer at the TiO$_2$/RuO$_2$ interface, we calculate the planar-averaged differential charge density,
\begin{equation}
\Delta\rho(z)=\rho_{\mathrm{TiO_2/RuO_2}}(z)
-\rho_{\mathrm{TiO_2}}(z)
-\rho_{\mathrm{RuO_2}}(z),
\end{equation}
where \(z\) is the coordinate normal to the interface, \(\rho_{\mathrm{TiO_2/RuO_2}}(z)\) is the planar-averaged electronic charge density of the full heterostructure, and \(\rho_{\mathrm{TiO_2}}(z)\) and \(\rho_{\mathrm{RuO_2}}(z)\) are those of isolated TiO$_2$ and RuO$_2$ slabs evaluated in the same atomic geometry. As shown in Fig.~\ref{fig:charge}(a), sizable charge redistribution is confined to the interfacial atomic layers and decays rapidly away from the interface, closely following the spatial profile of the induced Ru moments.
Integrating $\Delta\rho(z)$ over interfacial windows on the TiO$_2$ and RuO$_2$ sides yields $\Delta Q_{\mathrm{TiO_2}}<0$ and $\Delta Q_{\mathrm{RuO_2}}>0$ [see Fig.~\ref{fig:charge}(a) inset], indicating net electron transfer from TiO$_2$ toward interfacial RuO$_2$. Interesting, as discussed below, this charge transfer does not amount to simple electron doping of the Ru $4d$ orbitals.

To identify the orbital character of the electronic reconstruction, we compare the orbital-projected L\"owdin occupations of interfacial Ru and O atoms with the averages over the four central RuO$_2$ layers, which serve as a bulk-like reference. The interfacial Ru atom loses $0.045\,e$ from its $4d$ manifold, mainly due to a $0.082\,e$ depletion of the $d_{xy}$ orbital, while the other $d$ orbitals show small occupation increases [Fig.~\ref{fig:charge}(b)]. In contrast, the two neighboring O atoms gain $0.085\,e$ in their $2p$ manifolds, primarily in the $p_z$ orbitals [Fig.~\ref{fig:charge}(c)]. The net charge gain within one interfacial RuO$_2$ unit is therefore $0.040\,e$ [Fig.~\ref{fig:charge}(d)], consistent with the  differential charge density analysis. Thus, although electrons are transferred into the RuO$_2$ side of the interface, the added charge resides mainly on O ligand states, while the Ru $4d$ manifold is effectively hole doped.

This reduced Ru-$d$ filling provides a microscopic explanation for the magnetic instability. Previous studies showed that hole doping can induce antiferromagnetism in bulk RuO$_2$ under suitable tensile strain~\cite{Qian25p174425}. The TiO$_2$/RuO$_2$ interface realizes this tuning locally: epitaxial tensile strain modifies the Ru--O hopping network, while interfacial orbital rehybridization depletes the Ru-derived states, driving the boundary layers across the near-critical itinerant magnetic instability.

\begin{figure}[t]
  \centering
  \includegraphics[width=0.49\textwidth]{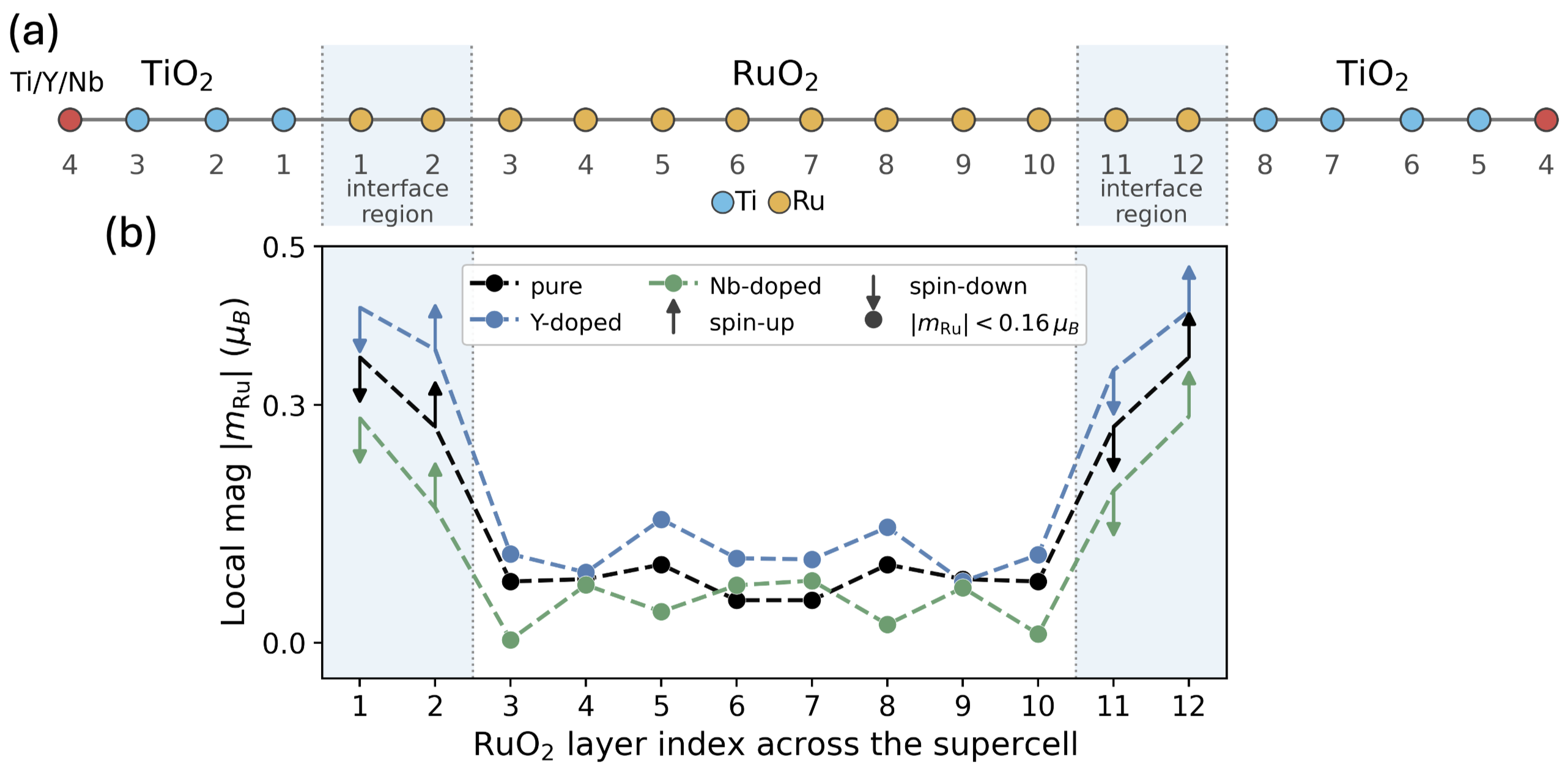}
  \caption{(a) Schematic of the pristine, Y-doped, and Nb-doped TiO$_2$/RuO$_2$/TiO$_2$ heterostructures. (b) Layer-resolved absolute Ru magnetic moments, $|m_{\mathrm{Ru}}|$, in pristine, Y-doped, and Nb-doped TiO$_2$/RuO$_2$ heterostructures. The $p$-type Y substitution in the central region of TiO$_2$ enhances the interfacial Ru moments, whereas $n$-type Nb substitution suppresses them.}
  \label{fig:doping}
\end{figure}

We next performed model calculations to test how interfacial charge transfer affects the induced magnetism. The charge redistribution was tuned by substituting Ti atoms in the central region of the TiO$_2$ substrate with $p$-type Y dopants or $n$-type Nb dopants, as illustrated in Fig.~\ref{fig:doping}(a). Figure~\ref{fig:doping}(b) shows that Y substitution enhances the interfacial Ru moments, whereas Nb substitution suppresses them. The corresponding charge analysis shows that Y reduces, while Nb increases, the net electron accumulation on the RuO$_2$ side.
This trend rules out a simple electron-accumulation mechanism: the larger electron transfer to RuO$_2$ in the Nb-doped case does not produce larger Ru moments. Instead, the result supports an orbital-selective mechanism. Y doping reinforces the depletion of Ru-$d$ states, whereas Nb doping partially compensates it. Thus, the interfacial magnetism is controlled by the Ru-$d$ filling rather than by the total transferred charge. This conclusion is consistent with the hole-enhanced magnetic instability previously identified in strained bulk RuO$_2$~\cite{Qian25p174425}.

\begin{figure}
  \centering
  \includegraphics[width=0.48\textwidth]{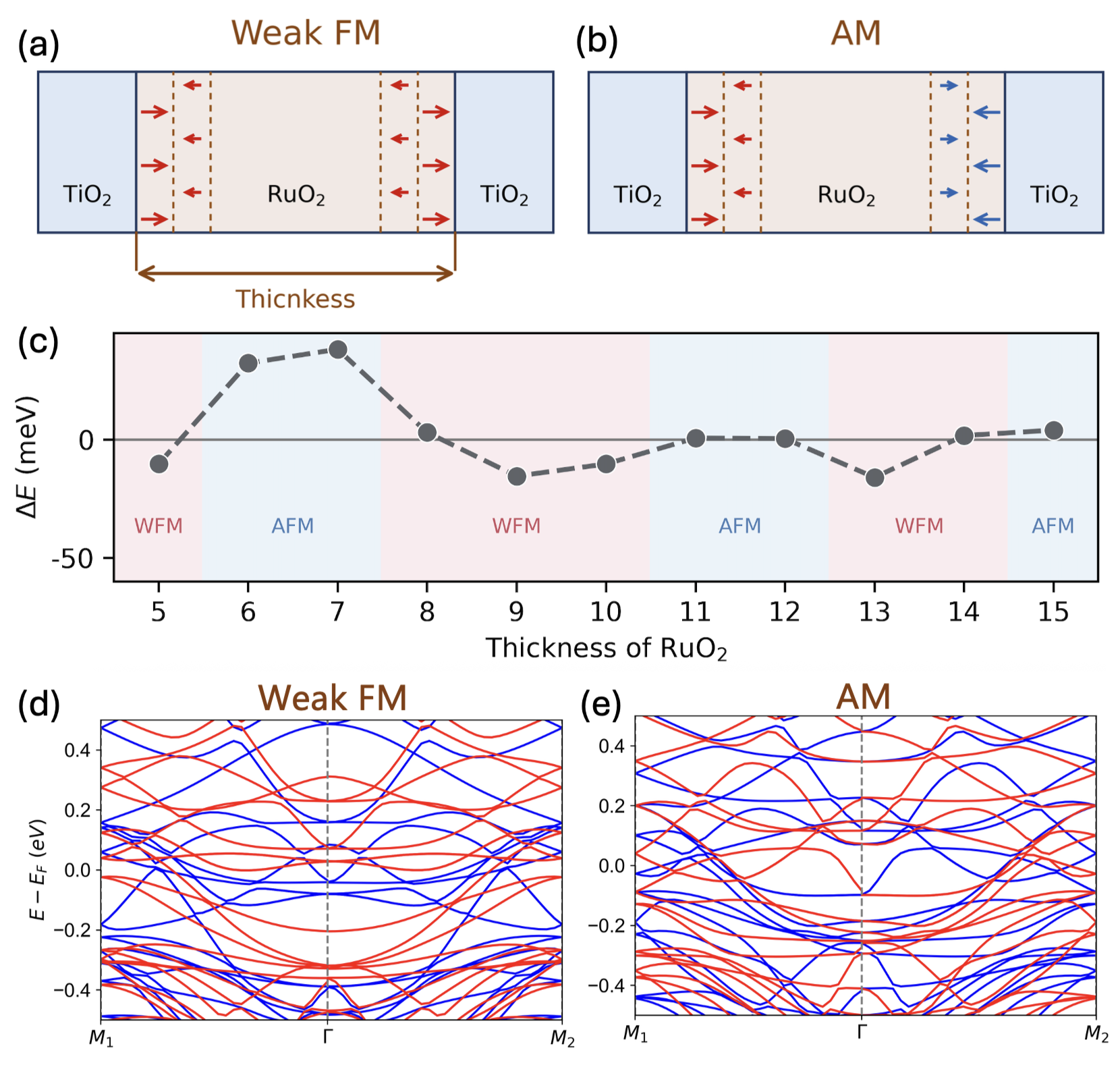}
  \caption{Interfacial magnetic coupling in TiO$_2$/RuO$_2$/TiO$_2$ sandwich structures.
(a) Parallel alignment of the interfacial moments produces a weak ferromagnetic (WFM) state with a finite net magnetization.
(b) Antiparallel alignment produces a fully compensated altermagnetic (AM) state. Both ordered states are lower in energy than the nonmagnetic state.
(c) Energy difference $\Delta E = E_{\mathrm{WFM}} - E_{\mathrm{AM}}$ as a function of RuO$_2$ thickness. Positive values favor the AM state, whereas negative values favor the WFM state. The oscillatory ground-state sequence indicates that the two magnetic boundary layers remain coupled through the RuO$_2$ interior via an RKKY-like itinerant exchange.
(d,e) Spin-resolved band structures of the WFM and AM states, respectively, for a six-layer RuO$_2$ slab sandwiched between TiO$_2$ layers..}
  \label{fig:coupling}
\end{figure}

As shown in Fig.~\ref{fig:interface}, spin polarization is confined mainly to the two Ru layers closest to each TiO$_2$/RuO$_2$ interface. The Ru layer directly adjacent to the interface carries the larger local moment, while the moment in the next Ru layer is smaller and rapidly decays toward the RuO$_2$ interior. These interfacial Ru moments form an antiferromagnetic arrangement within each boundary region; however, because the two spin-polarized layers have unequal moment sizes, each interface retains an uncompensated net magnetic moment. This behavior suggests an interesting scenario: in a symmetric TiO$_2$/RuO$_2$/TiO$_2$ heterostructure, the two magnetic interfaces may couple through the metallic RuO$_2$ spacer and align either parallel or antiparallel, as schematically illustrated in Fig.~\ref{fig:coupling}. Parallel alignment of the uncompensated interfacial moments produces a finite total magnetization, which we denote as the weak-ferromagnetic (WFM) state [Fig.~\ref{fig:coupling}(a)]. In contrast, antiparallel alignment cancels the net moments of the two interfaces, giving a fully compensated antiferromagnetic configuration. By symmetry, this compensated state retains the altermagnetic character of RuO$_2$, and we therefore refer to it as the AM state [Fig.~\ref{fig:coupling}(b)].

To determine which configuration is favored, we calculate the total-energy difference
$\Delta E = E_{\mathrm{WFM}}-E_{\mathrm{AM}}$,
as a function of the RuO$_2$ thickness. As shown in Fig.~\ref{fig:coupling}(c), \(\Delta E\) exhibits an oscillatory sign change with increasing RuO$_2$ thickness, indicating that the magnetic ground state alternates between WFM and AM configurations. This behavior is reminiscent of RKKY-like interlayer exchange coupling, in which two magnetic regions interact through the itinerant electrons of a metallic spacer. Although the static local moments are localized near the TiO$_2$/RuO$_2$ interfaces, the metallic RuO$_2$ interior can mediate a thickness-dependent exchange interaction between them.

We further examined the electronic band structures for the six-layer RuO$_2$ slab, for which the AM state is lower in energy than the WFM state. As shown in Fig.~\ref{fig:coupling}(d), the WFM state exhibits the conventional spin splitting expected for a ferromagnetic metal. In contrast, the AM state remains metallic and shows momentum-dependent spin splitting characteristic of altermagnetism [Fig.~\ref{fig:coupling}(e)], while retaining zero net magnetization.
These results show that TiO$_2$/RuO$_2$/TiO$_2$ heterostructures provide a tunable platform for engineering magnetic states in RuO$_2$. By varying the RuO$_2$ thickness, one can select between WFM and AM ground states; by modifying the interfacial charge environment through doping of the TiO$_2$ layers, one can further tune the induced magnetic moments. Future work could explore how the periodicity of TiO$_2$/RuO$_2$ superlattices controls the interfacial coupling and the resulting magnetic phases.

In summary, we have shown that the magnetism observed in epitaxial RuO$_2$ can be understood as an emergent interfacial boundary phase. While TiO$_2$-imposed in-plane strain alone does not stabilize magnetism in (001)-oriented RuO$_2$, explicit TiO$_2$/RuO$_2$ interfaces generate sizable Ru local moments confined to the first few boundary layers. The absence of comparable moments at a vacuum-terminated RuO$_2$ surface, together with the disappearance of interfacial magnetism when the heterostructure is constrained to the bulk RuO$_2$ in-plane lattice constants, highlights the cooperative role of epitaxial strain and interfacial electronic reconstruction. Charge-density analysis further reveals localized charge redistribution from TiO$_2$ toward interfacial RuO$_2$.
Importantly, orbital-selective hole doping of the Ru $4d$ orbitals is responsible for the emergence of local magnetic moments.
By tuning the substrate with Y or Nb dopants, the interfacial charge environment and the resulting Ru moments can be systematically modulated. Finally, in symmetric TiO$_2$/RuO$_2$/TiO$_2$ heterostructures, the two magnetic interfaces remain coupled through the metallic RuO$_2$ spacer, leading to an RKKY-like thickness-dependent alternation between weak-ferromagnetic and fully compensated altermagnetic ground states.

These findings establish TiO$_2$/RuO$_2$ heterostructures as a tunable platform for controlling fragile magnetism in near-critical itinerant metals. In particular, the RuO$_2$ thickness provides a direct handle for selecting between uncompensated and compensated magnetic boundary states, while substrate doping offers an additional route to tune the strength of the interfacial moments. More broadly, our results suggest that nonmagnetic oxide boundaries can be used to impose strain as well as to engineer electronic reconstructions to control magnetic order. Future work exploring different interface terminations, dopant distributions, superlattice periodicities, and electrostatic gating could further expand this strategy, enabling on-demand design of altermagnetic and weak-ferromagnetic phases for spintronic applications.

\begin{acknowledgments}
C. W. acknowledges support from the National Natural Science Foundation of China (Grants No. 12234016 and No. 12174317), and S. L. from the Zhejiang Provincial Natural Science Foundation of China (Grant No. LR25A040004). This work was supported by the New Cornerstone Science Foundation. Computational resources were supported by the Open Source Supercomputing Center of S-A-I.
\end{acknowledgments}

\bibliography{note}
\end{document}